\documentclass[twocolumn,prd,showpacs,nofootinbib]{revtex4}

\usepackage{graphicx}
\usepackage{dcolumn}
\usepackage{bm}

\def\VEV#1{\left\langle #1 \right\rangle}
\def\Mpl{M_{\mathrm{ Pl}}}

\newcommand{\beq}{\begin{equation}}
\newcommand{\eeq}{\end{equation}}
\newcommand{\beqa}{\begin{eqnarray}}
\newcommand{\eeqa}{\end{eqnarray}}

\newcommand{\lcdm}{$\Lambda$CDM }

\newcommand{\xdec}{x_\mathrm{d}}
\newcommand{\zdec}{z_\mathrm{d}}
\newcommand{\taudec}{\tau_\mathrm{d}}

\begin{document}

\title{A Hemispherical Power Asymmetry from Inflation}

\author{Adrienne L.~Erickcek, Marc Kamionkowski, and Sean M.~Carroll}
\affiliation{California Institute of Technology, Mail Code 130-33,
Pasadena, CA 91125}

\date{\today}

\begin{abstract}
Measurements of CMB temperature fluctuations by the Wilkinson Microwave Anisotropy Probe (WMAP) indicate that the fluctuation amplitude in one half of the sky differs from the amplitude in the other half.  We show that such an asymmetry cannot be generated during single-field slow-roll inflation without violating constraints to the homogeneity of the Universe.  In contrast, a multi-field inflationary theory, the curvaton model, can produce this power asymmetry without violating the homogeneity constraint.  The mechanism requires the introduction of a large-amplitude superhorizon perturbation to the curvaton field, possibly a pre-inflationary remnant or a superhorizon curvaton-web structure.  The model makes several predictions, including non-Gaussianity and modifications to the inflationary consistency relation, that will be tested with forthcoming CMB experiments.
\end{abstract}


\maketitle

\section{Introduction}

Inflation provides a compelling description of the early
Universe \cite{Guth:1980zm}.  The
temperature fluctuations in the cosmic microwave background
(CMB) \cite{deBernardis:2000gy,Dunkley:2008ie} and the
distribution of galaxies \cite{Cole:2005sx} agree
well with inflationary predictions.  However, there is an
anomaly in the CMB: measurements from the Wilkinson Microwave
Anisotropy Probe (WMAP) \cite{Dunkley:2008ie} indicate that the
temperature-fluctuation amplitude is larger, by roughly 10\%, in
one hemisphere than in the other \cite{Eriksen:2003db}.  Fewer than 1\%
of simulated isotropic fluctuation maps exhibit such an
asymmetry, and the asymmetry cannot be
attributed to any known astrophysical foreground or experimental
artifact.  As opposed to the ``axis of evil'' \cite{deOliveiraCosta:2003pu}, an apparent alignment of only the lowest multipole moments, this asymmetry has gone largely unnoticed (although see \cite{Gordon:2006ag,Donoghue:2007ze}), and it warrants further theoretical consideration.

\begin{figure}[b]
\includegraphics[width=8.5cm]{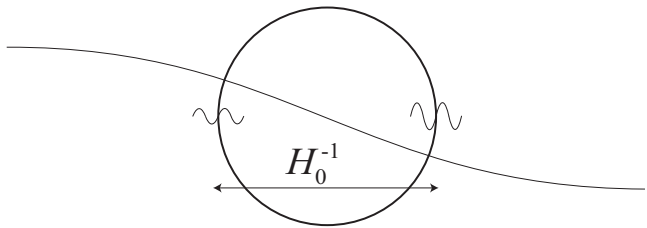}
\caption{Measurements of temperature fluctuations in the
     cosmic microwave background (CMB) show that the rms
     temperature-fluctuation amplitude is larger in one side of
     the sky than in the other.  We investigate here whether
     this may arise as a consequence of a large-scale mode of an
     inflaton or curvaton.}
\label{fig:supermode}
\end{figure}

In the standard inflation scenario, the Universe undergoes a very long inflationary
expansion before the comoving observable Universe exits the horizon during
inflation.  Thus, any remnants of a pre-inflationary Universe were inflated
away before there could be observable consequences.  This accounts for the
smoothness of the primordial Universe as well as its flatness.  It also
suggests that primordial density perturbations should show no preferred
direction.  The existence of a hemispherical power asymmetry in the CMB challenges this basic prediction of inflation.

A superhorizon perturbation would introduce a preferred direction in the Universe and has been considered as a possible origin of the ``axis of evil" \cite{Gordon:2005ai}.    In this article, we investigate how the hemispherical power asymmetry could result from a superhorizon perturbation during inflation,  as depicted in Fig.~\ref{fig:supermode}.  Since the amplitude of the primordial fluctuations depends on the background value of the fluctuating inflationary field, a large-amplitude superhorizon fluctuation would generate a power asymmetry by varying the background value of the field across the observable Universe.  Of course, the superhorizon fluctuation would make the Universe inhomogeneous, and the near-uniformity of the CMB constrains such departures from homogeneity \cite{Erickcek:2008jp}.   

We begin by showing in section \ref{sec:singlefield} that the power asymmetry cannot be reconciled with single-field slow-roll inflation without violating constraints to the homogeneity of the Universe.   We then consider an alternative inflationary theory, the curvaton model \cite{Moroi:2001ct}, which has been suggested as a possible source of a power asymmetry \cite{Gordon:2006ag}.  In section \ref{sec:curvaton}, we demonstrate that a superhorizon fluctuation in the curvaton field can generate the observed asymmetry without violating the homogeneity constraints.   The required superhorizon fluctuation in the curvaton field may occur, for example, as a remnant of the pre-inflationary epoch or as a signature of superhorizon curvaton-web structures \cite{Linde:2005yw}.  The proposed model predicts several signatures, which may soon be tested, in the CMB.  We discuss these signatures and summarize our findings in section \ref{sec:discussion}.

\section{Single-Field Models}
\label{sec:singlefield}

Inflation postulates that the energy
density in the early Universe was
dominated by a scalar field $\phi$, the inflaton.  The energy density is due to
kinetic energy $(1/2)\dot\phi^2$ plus some potential energy $V(\phi)$.  If the
slow-roll parameters, $\epsilon \equiv (\Mpl^2/16\pi)(V'/V)^2$ and $\eta \equiv
(\Mpl^2/8\pi)(V''/V)$, are small, then the field rolls slowly.  The energy
density is then dominated by the potential energy, the pressure is negative,
and the expansion of the Universe is inflationary.

Quantum fluctuations in the inflaton give rise to primordial density
perturbations characterized by a gravitational-potential power spectrum
$P_\Phi(k) \propto V/\epsilon$, where $V$ and $\epsilon$ are evaluated at the
value the inflaton took when the comoving wavenumber $k$
exited the horizon during inflation. Differentiation of the expression for
$P_\Phi(k)$ suggests that the power spectrum can be approximated as $P_\Phi(k)
\propto k^{n_s-1}$, where the scalar spectral index $n_s=1-6\epsilon + 2\eta$
is close to unity, consistent with current
measurements \cite{Dunkley:2008ie,Verde:2008ix}.

The power spectrum $P_\Phi(k)$ may vary with $k$
because different values of $k$ sample the quantity $V/\epsilon$
at different values of the inflaton $\phi$.  This suggests that
the power asymmetry might be explained by a large-amplitude mode
of $\phi$ with comoving wavelength long compared with the
current Hubble distance ($k\ll H_0$).  Then one side of the CMB sky would
reflect the imprint of a different value of $\phi$ than the other side.
{}From $P_\Phi(k) \propto V/\epsilon$, we infer a fractional power asymmetry,
\begin{equation}
 A\equiv \frac{\Delta P_\Phi}{P_\Phi}=-2\sqrt{\frac{\pi}{\epsilon} }(1-n_s) \frac{\Delta \phi}{\Mpl},
 \label{inflationA}
 \end{equation}
where $\Delta\phi$ is the change in the inflaton field across the observable
Universe.   A 10\% variation in the amplitude of the CMB temperature fluctuations corresponds to a power asymmetry $A=0.2$.

The gravitational-potential perturbation $\Phi$ during matter
domination is related to the inflaton perturbation $\delta \phi$
through $\Phi = (6/5)\sqrt{\pi/\epsilon}(\delta\phi / \Mpl)$.
Thus, a long-wavelength perturbation $\delta\phi \propto \sin
[\vec{k}\cdot\vec{x}+\varpi]$, with $k\xdec\ll1$ (where
$\xdec$ is the distance to the surface of last scatter), introduces
a gravitational-potential perturbation with the same spatial
dependence.  It follows from Eq.~(\ref{inflationA}) that $\Delta
\Phi = 3A/[5(n_s-1)]$.  An immediate concern, therefore, is
whether this large-amplitude perturbation is consistent with the
isotropy of the CMB.

Gravitational-potential perturbations give rise to temperature
fluctuations in the CMB through the Sachs-Wolfe effect
\cite{Sachs:1967er} ($\delta T/T \simeq \Phi/3$).  A large-scale potential perturbation might thus
be expected to produce a CMB temperature dipole of similar
magnitude.  However, for the Einstein-de Sitter universe, the
potential perturbation induces a peculiar velocity whose Doppler
shift cancels the intrinsic temperature dipole
\cite{GZ}.  The same is true for a flat universe
with a cosmological constant \cite{Erickcek:2008jp}.

Although the dipole vanishes, measurements of the CMB
temperature quadrupole and octupole constrain the cosmological
potential gradient \cite{GZ,Castro:2003bk}.  Here
we outline how these constraints are derived; the
full calculation is presented elsewhere
\cite{Erickcek:2008jp}.   Since $k \xdec \ll 1$, we first
expand the sinusoidal dependence $\Phi(\vec x) = \Phi_{\vec
k} \sin(\vec k \cdot \vec x +\varpi)$ in powers of $\vec
k \cdot \vec x$.  Then the terms that contribute to the CMB
quadrupole and octupole are
\beq
     \Phi(\vec{x})=-\Phi_{\vec k}\left\{
        \frac{(\vec{k}\cdot\vec{x})^2}{2} \sin \varpi
      +\frac{(\vec{k}\cdot\vec{x})^3}{6}\cos \varpi \right\}.
\label{PhiExp}
\eeq
The CMB temperature anisotropy produced by the
potential in Eq.~(\ref{PhiExp}) is
\beqa
     \frac{\Delta T}{T}(\hat{n}) = - \Phi_{\vec k}\left[
     \frac{\mu^2}{2} (k\xdec)^2 \delta_2 \sin \varpi +
     \frac{\mu^3}{6} (k\xdec)^3 \delta_3\cos \varpi \right],
\label{delT}
\eeqa
where $\mu \equiv \hat{k}\cdot\hat{n}$ and $\Phi_{\vec k}$ is evaluated at the time of decoupling ($\taudec$). The $\delta_i$ account for the Sachs-Wolfe (including integrated) effect and the Doppler effect induced by $\Phi_{\vec k}$; for a \lcdm Universe with $\Omega_M = 0.28$ and decoupling redshift
$\zdec = 1090$, we find that $\delta_2=0.33$ and $\delta_3=0.35$.  Choosing
$\hat{k} = \hat{z}$,
Eq.~(\ref{delT}) gives nonzero values for the spherical-harmonic
coefficients $a_{20}$ and $a_{30}$.  The relevant observational
constraints are therefore,
\beqa
     (k \xdec)^2 \left|\Phi_{\vec k}(\taudec)\sin \varpi  \right|
     &\lesssim& 5.8 \, Q \label{quad}\\
     (k \xdec)^3 \left|\Phi_{\vec k}(\taudec)\cos \varpi \right|
     &\lesssim& 32 \,  {\cal O} \label{oct}
\eeqa
where $Q$ and ${\cal O}$ are upper bounds on $|a_{20}|$ and
$|a_{30}|$, respectively, in a coordinate system aligned
with the power asymmetry.  
We take $Q = 3\sqrt{C_2} \lesssim 1.8\times10^{-5}$ and ${\cal
O} = 3\sqrt{C_3} \lesssim 2.7 \times 10^{-5}$, 3 times the
measured rms values of the quadrupole and octupole
\cite{Efstathiou:2003tv}, as $3\sigma$ upper limits; this
accounts for cosmic variance in the quadrupole and octupole due
to smaller-scale modes.   The temperature
quadrupole and octupole induced by the superhorizon mode can be made arbitrarily small for fixed
$\Delta \Phi \simeq \Phi_{\vec k}  (k \xdec) \cos \varpi$ by choosing
$k$ to be sufficiently small.   However, we also demand that $\Phi_{\vec k}
\lesssim 1$ everywhere, and this sets a lower bound on $(k \xdec)$.

We now return to the power asymmetry generated by an inflaton
perturbation.  The largest value of $\Delta \Phi$ is
obtained if $\varpi=0$, in which case the perturbation produces
no quadrupole.  The octupole constraint [Eq.~(\ref{oct})]
combined with $(k \xdec) \gtrsim |\Delta \Phi|$ [i.e., the
requirement $\Phi_{\vec k} \lesssim 1$] implies that $|\Delta \Phi|
\lesssim (32 \, {\cal O})^{1/3}$.  Given that $(1-n_s)
\lesssim 0.06$, we see 
that the maximum possible power asymmetry obtainable with a
single superhorizon mode is  $A_{\mathrm{ max}}\simeq 0.1(32\,
{\cal O})^{1/3} \simeq 0.0095$.  This is too small, by
more than an
order of magnitude, to account for the observed asymmetry.  The
limit can be circumvented if a number of Fourier modes conspire
to make the density gradient across the observable Universe
smoother.  This would require, however, that we live in a very
special place in a very unusual density distribution.

\section{The Curvaton Model}
\label{sec:curvaton}

We thus turn our attention to the curvaton model
\cite{Moroi:2001ct}
of inflation. This model introduces a second scalar field
$\sigma$, the curvaton, with potential $V(\sigma)= (1/2)
m_\sigma^2 \sigma^2$.  During inflation, it is effectively
massless, $m_\sigma \ll H_I$ (where $H_I$ is the inflationary
expansion rate), and its density is negligible.  Its homogeneous
value $\bar \sigma$ remains classically frozen during inflation,
but quantum effects give rise to fluctuations $\delta\sigma$ of
rms amplitude $(\delta \sigma)_{\mathrm{ rms}} \simeq
(H_I/2\pi)$. Well after inflation, the curvaton rolls
toward its minimum and then later oscillates about its
minimum---i.e., a cold gas of $\sigma$ particles.  These
particles then decay to radiation.  The fluctuations in the
curvaton field will produce gravitational-potential
perturbations with power spectrum,
\beq
P_{\Phi,\sigma} \propto
    R^2 \VEV{ \left[\frac{\delta V}{V(\bar \sigma)}\right]^2 }
    \sim R^2\left( \frac{H_I}{\pi \bar \sigma}
     \right)^2, 
\eeq
provided that $\bar \sigma \gg H_I$ \cite{Malik:2002jb}.  Here $R\equiv(\rho_\sigma/\rho_\mathrm{tot})$
is the energy density of the curvaton field just prior to its decay divided by the total energy density of the Universe at that time.  

We hypothesize that the density due to
curvaton decay is small compared with the density 
due to inflaton decay; i.e., $R \ll 1$.
In this case, the perturbation in the total energy density, and
thus the potential perturbation $\Phi$, due to a fluctuation in
$\rho_\sigma$ will be suppressed, making it possible to satisfy
the homogeneity conditions set by the CMB [Eqs.~(\ref{quad}) and
(\ref{oct})], even if $\rho_\sigma$ has order-unity variations.
We then hypothesize that the power asymmetry comes from a
variation $\Delta\bar\sigma$ in the value of the mean curvaton field 
across the observable Universe.  Since $R\propto \bar \sigma^2$ for $R\ll 1$, the power spectrum for gravitational-potential perturbations produced by the curvaton is proportional to $\bar\sigma^2$.
A variation $\Delta\bar\sigma$ in the value of the mean curvaton field 
across the observable Universe therefore induces a fractional power asymmetry $\Delta
P_{\Phi,\sigma}/P_{\Phi,\sigma} \simeq 2(\Delta \bar
\sigma/\bar \sigma)$.

First we must ensure that this inhomogeneity does not violate
Eqs.~(\ref{quad}) and (\ref{oct}).  The potential fluctuation
during matter domination produced by a fluctuation $\delta
\sigma$ in the curvaton field is
\beq
     \Phi= - \frac{R}{5}\left[2
     \left(\frac{\delta\sigma}{\bar\sigma}\right) +
     \left(\frac{\delta\sigma}{\bar\sigma}\right)^2
     \right].
\label{sigmaPhi}
\eeq  
Consider a superhorizon sinusoidal perturbation to the
curvaton field $\delta\bar\sigma = \sigma_k
\sin(\vec{k}\cdot\vec{x}+\varpi)$.  If we ignored the
term in Eq.~(\ref{sigmaPhi}) quadratic in $\delta\sigma$, then
the upper bound to $\delta\bar\sigma$ would be obtained by
setting $\varpi=0$.  As with the inflaton, the constraint would then
then arise from the CMB octupole.  However, the term in
Eq.~(\ref{sigmaPhi}) quadratic in $\delta\sigma$ gives rise to a
term in $\Phi$ quadratic in $(\vec k\cdot \vec x)$---i.e.,
$\Phi_{\mathrm{quad}} = -(R/5)(\sigma_k/\bar\sigma)^2 (\vec k
\cdot \vec x)^2$ for $\varpi=0$.  Noting that $(\Delta\bar\sigma/\bar\sigma) =
(\sigma_k/\bar\sigma)(\vec k \cdot\vec \xdec)$, the quadrupole bound in
Eq.~(\ref{quad}) yields an upper limit,
\beq
     R\left( \frac{\Delta\bar\sigma}{\bar\sigma}\right)^2
     \lesssim \frac{5}{2}(5.8\,Q). 
\label{curvatonQ}
\eeq
While this bound was derived for $\varpi=0$, most other values for $\varpi$ yield similar constraints \cite{Erickcek:2008jp}.

\begin{figure}
\includegraphics[width=8.5cm]{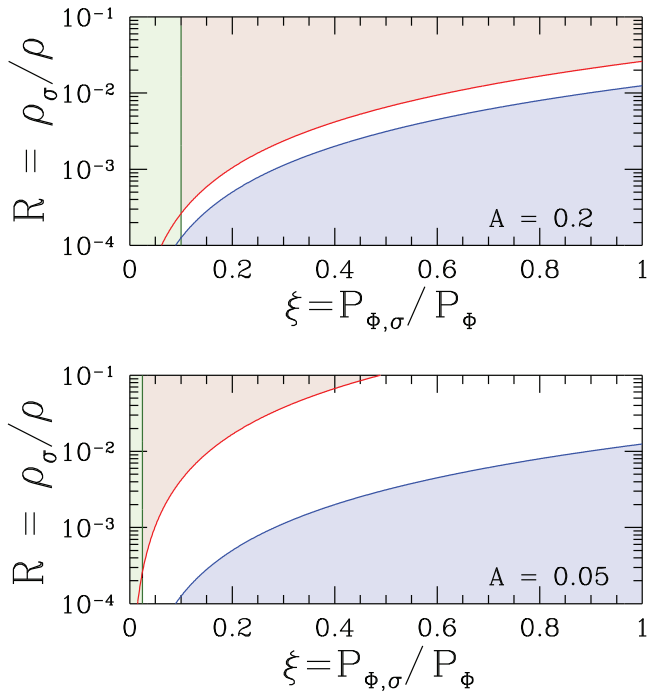}
\caption{The $R$-$\xi$ parameter space for the
     curvaton model that produces a power asymmetry $A=0.2$ (top) and $A=0.05$ (bottom).
     Here $R$ is the fraction of the cosmological density due to
     curvaton decay, and $\xi$ is the fraction of the power due
     to the curvaton.  The upper limit to $R$ comes from the
     CMB-quadrupole constraint.  The lower bound comes from $f_{\mathrm
     NL}\leq100$.  The lower limit to $\xi$ comes from the requirement
     that the fractional change in the curvaton field across the
     observable Universe be less than one.  If $A$ is lowered,
     the lower bound to $R$ remains unchanged, but the upper bound increases,
     proportional to $A^{-2}$.  The lower limit to $\xi$ also
     decreases as $A$ decreases, proportional to $A$.}
\label{fig:plot}
\end{figure}

Most generally, the primordial power will be some combination of
that due to the inflaton and curvaton \cite{Langlois:2004nn}, $P_\Phi=P_{\Phi,\phi}
+P_{\Phi,\sigma} \simeq 10^{-9}$, with a fraction $\xi \equiv
P_{\Phi,\sigma}/P_\Phi$ due to the curvaton.   The required
asymmetry, $A \simeq 2\xi (\Delta \bar\sigma/\bar \sigma)$, can be
obtained without violating Eq.~(\ref{curvatonQ}) by choosing $R
\lesssim 58\, Q \xi^2/A^2$, as shown in Fig.~\ref{fig:plot}.

The only remaining issue is the Gaussianity of primordial
perturbations.  The curvaton fluctuation $\delta\sigma$ is a
Gaussian random variable.  Since the curvaton-induced density
perturbation has a contribution quadratic in $(\delta\sigma)^2$,
it implies a non-Gaussian contribution to the density
fluctuation. The departure from
Gaussianity can be estimated from the parameter
$f_{\mathrm{NL}}$ \cite{Verde:1999ij}, which
for the curvaton model is $f_{\mathrm{ NL}} \simeq 5\xi^2/(4R)$
\cite{Lyth:2002my, Malik:2006pm, Ichikawa:2008iq}.  The current upper limit,
$f_{\mathrm{NL}} \lesssim 100$
\cite{Komatsu:2003fd}, leads
to the lower limit to $R$ shown in Fig.~\ref{fig:plot}.

Fig.~\ref{fig:plot} shows that there are values of $R$ and $\xi$
that lead to a power asymmetry $A=0.2$ and are consistent with
measurements of the CMB quadrupole and $f_{\mathrm{NL}}$.  For
any value of $A$, the allowed region of $R$-$\xi$ parameter
space is
\beq
     \frac{5}{4 \, f_{\mathrm{NL},\mathrm{max}}}\lesssim
     \frac{R}{\xi^2} \lesssim 58 \frac{Q}{A^2},
\eeq
where $f_{\mathrm{NL},\mathrm{max}}$ is the largest allowed
value for $f_{\mathrm{NL}}$.   Thus, we see that measurements
of the CMB quadrupole and $f_\mathrm{NL}$ place an upper bound,
\beq
     A \lesssim \sqrt{(58 \, Q)\,\left(\frac{4
     f_{\mathrm{NL},\mathrm{max}}}{5}\right) },
\eeq
on the power asymmetry that may be generated by a superhorizon
curvaton fluctuation.
For $Q = 1.8\times10^{-5}$, we predict (for
$A\simeq0.2$) $f_{\mathrm{NL}} \gtrsim 50$, much
larger than $f_{\mathrm{NL}}\ll 1$ predicted by
standard slow-roll inflation.  Values as small as
$f_{\mathrm{NL}}\simeq 5$ should be accessible to the
forthcoming Planck satellite, and so there should be a clear
signature in Planck if the power asymmetry was generated by a curvaton perturbation and $A=0.2$.

If $(\delta\sigma/\bar\sigma) \ll1$, the power due to
the curvaton is $P_{\Phi,\sigma} \simeq
(2R/5)^2\VEV{(\delta\sigma/\bar\sigma)^2}$.
The power required from the curvaton fixes
$R(\delta\sigma/\bar\sigma)_{\mathrm{rms}} \simeq
8\times10^{-5}\, \xi^{1/2}$, from which it follows that
$(\delta\sigma/\bar\sigma)_{\mathrm{rms}} \lesssim 0.2$ for the
allowed parameter space in Fig.~\ref{fig:plot}, thus verifying
that this parameter is small.  We find from
$(\Delta\bar\sigma/\bar\sigma) = A/2\xi \lesssim 1$ that the required
cross-horizon variation $\Delta\bar\sigma/\bar\sigma$ in the
curvaton is large compared with the characteristic quantum-mechanical curvaton fluctuation
$(\delta\sigma/\bar\sigma)_{\mathrm{rms}}$;
the required $\Delta\bar\sigma$ is at least a
$\sim5\sigma$ fluctuation.  It may therefore be that this
large-scale mode is a superhorizon inhomogeneity not completely
erased by inflation.  Another possibility is that 
positive- and negative-value cells of $\bar \sigma$ created
during inflation may be large enough to encompass the observable
Universe; if so, we would observe an order-unity fluctuation in
$\bar \sigma$ near the $\bar \sigma =0$ wall that divides two
cells \cite{Linde:2005yw}.

\section{Summary and Discussion}
\label{sec:discussion}

The hemispherical power asymmetry in the CMB challenges the assumption that the Universe is isotropic and homogeneous.  A superhorizon perturbation in an inflationary field would introduce a preferred direction in the Universe, and we have investigated this mechanism for generating the observed power asymmetry.  We found that the required superhorizon fluctuation in the inflaton field is inconsistent with measurements of the CMB octupole.  A superhorizon fluctuation in a subdominant scalar field, however, is a viable alternative.  A superhorizon curvaton perturbation can generate the observed power asymmetry without introducing unacceptable anisotropy and non-Gaussianity in the CMB.  

We have considered the specific asymmetry $A\simeq0.2$
reported for WMAP, but our results can be scaled for
different values of $A$, should the measured value for the
asymmetry change in the future.  In particular, the
$f_{\mathrm{NL}}$ constraint (the lower bound to $R$) in
Fig.~\ref{fig:plot} remains the same, but the upper bound (from
the quadrupole) increases as $A$ is decreased.  The
lower limit to $\xi$ also decreases as $A$ is
decreased. Here we have also considered
a general model in which primordial perturbations come from some
combination of the inflaton and curvaton.  Although it may seem
unnatural to expect the two field decays to produce comparable
fluctuation amplitudes, our mechanism works even if
$\xi=1$ (the fluctuations are due entirely to the curvaton).  Thus,
the coincidence is not a requirement of the model.

If the power asymmetry can indeed be attributed to a superhorizon curvaton
mode, then the workings of inflation are more subtle than the simplest models
would suggest.  Fortunately, the theory makes a number of
predictions that can be pursued with future experiments.  To
begin, the modulated power should produce signatures in the
CMB polarization and temperature-polarization correlations
\cite{Pullen:2007tu}. The curvaton model predicts
non-Gaussianity, of amplitude $f_{\mathrm{NL}}
\gtrsim50$ for $A\simeq0.2$, which will soon be experimentally
accessible.  However, the theory also predicts that the small-scale
non-Gaussianity will be modulated across the sky by the
variation in $\bar\sigma$ (and thus in $\xi$ and $R$).
The presence of curvaton fluctuations also changes other features of the
CMB \cite{Ichikawa:2008iq}.  The ratio of tensor and scalar
perturbations ($r$) is reduced by a factor of $(1-\xi)$
and the scalar spectral index is $n_s =
1-2\epsilon-(1-\xi)(4\epsilon-2\eta)$.  The tensor spectral
index ($n_T$), however, is unaltered by the presence of the
curvaton, and so this model alters the inflationary consistency
relation between $n_T$ and $r$ and possibly the prospects for
testing it \cite{Smith:2006xf}.

We have here assumed
simply that the curvaton decays to the same mixture of baryons,
dark matter, and radiation as the inflaton.  However, if the inflaton and
curvaton decays products differ, then there may be an
isocurvature component \cite{Lyth:2002my, Lemoine:2006sc}.  Finally, the
simplest scenario predicts a scale-invariant power asymmetry; 
the asymmetry has been found at multipole moments $\ell \lesssim 40$, 
but there are claims that it does not extend to higher $\ell$ \cite{Donoghue:2004gu}.  
If this result holds, it will be interesting to see whether the
departure from scale invariance can be obtained by suitably
altering the power spectra for the curvaton and inflaton.  For instance, a sudden drop 
in both $V^\prime(\phi)$ and $V(\phi)$ could enhance the gravitational-potential fluctuations from the inflaton while suppressing the fluctuations from the curvaton \cite{Gordon:2006ag}; the resulting drop in $\xi$ would reduce the power asymmetry on smaller scales.  We leave such elaborations for future work.

\acknowledgments

We thank K.~G\'orski and H.~K.~Eriksen for discussions.  This
work was supported by DoE DE-FG03-92-ER40701 and the Gordon and
Betty Moore Foundation.

\end{document}